# Enhancing heat transfer in X-ray tube by van der heterostructures-based thermionic emission

Sunchao Huang[1], Suguo Chen[2], Yue Wang[2], Xihang Shi[3], Xiaoqiuyan Zhang[1], Min Hu[1], Ping Zhang[1], Shaomeng Wang[1], Chao Zhang[4] and Yubin Gong[1†]

[1]School of Electronic Science and Engineering, University of Electronic Science and Technology of China, Chengdu, Sichuan 610054, China
[2]Department of Applied Physics, Xian University of Technology, Xian 710048, China
[3]Solid State Institute and Faculty of Electrical and Computer Engineering, Technion – Israel Institute of Technology, Haifa 32000, Israel
[4]School of Physics and Institute for Superconducting and Electronic Materials, University of Wollongong, Wollongong, NSW 2522, Australia
[†]Email: ybgong@uestc.edu.cn

**Abstract:** Van der Waals (vdW) heterostructures have attracted much attention due to their distinctive optical, electrical, and thermal properties, demonstrating promising potential in areas such as photocatalysis, ultrafast photonics, and free electron radiation devices. Particularly, they are promising platforms for studying thermionic emission. Here, we illustrate that using vdW heterostructure-based thermionic emission can enhance heat transfer in vacuum devices. As a proof of concept, we demonstrate that this approach offers a promising solution to the long-standing overheating issue in X-ray tubes. Specifically, we show that the saturated target temperature of a 2000 W X-ray tube can be reduced from around 1200°C to 490°C. Additionally, our study demonstrates that by reducing the height of the Schottky barrier formed in the vdW heterostructures, the thermionic cooling performance can be enhanced. Our findings pave the way for the development of high-power X-ray tubes.

Since the ground-breaking discovery of graphene in 2004[1], a large group of 2D van der Waals (vdW) materials have been uncovered. These materials exhibit exceptional properties, including linear energy-momentum dispersion[2], quantum Hall effects[3,4], strong light-matter interaction[5–8] and dangling-bond free surfaces[9]. The presence of dangling bond-free surfaces enables vdW materials to be stacked to form diverse vdW heterostructures without being limited by lattice mismatching[9]. VdW heterostructures have demonstrated promising applications in photocatalysis[10], ultrafast photonics[11], and free electron radiation devices[12]. Recently, vdW heterostructures have demonstrated excellent thermionic emission properties, holding promising applications in detecting and harvesting low-energy photons[13]. The excellent thermionic emission properties are attributed to the formation of a Schottky barrier within the vdW heterostructures[13]. However, to the best of our knowledge, the application of vdW heterostructures in solving the long-standing overheating issue of X-ray tubes has not been explored.

X-ray tubes are a kind of vacuum devices that convert electrical energy into X-rays via bremsstrahlung radiation and characteristic radiation. X-ray tubes are extensively utilized in various fields, including security screening, chip inspection, X-ray crystallography, and material analysis, owing to their compact design and cost-effectiveness[14–16]. Nevertheless, X-ray tubes face a persistent issue of overheating over prolonged periods, primarily due to their low energy conversion efficiency[16]. For instance, only 1% of the input electrical energy is transformed into X-rays, while the remaining energy is converted into heat within the target[16]. Despite the proposal of various methods, such as rotatable target design[17] and water-cooling technology[18], to address the overheating issue, the advancement of high-power X-ray tubes continues to be constrained by this problem. Lately, thermionic emission from Dirac materials has emerged as a promising solution to tackle this issue[19]. Thermionic emission refers to the release of electrons from an electrode due to its temperature, a process that takes place when the thermal energy imparted to the charge carrier surpasses the material's work function[20–25]. This process has been extensively studied for its potential applications in electricity generation[26–30] and thermionic cooling[31–37]. To considerably lower the saturated target temperature of an X-ray tube, the work function of the thermionic materials needs to be approximately 0.5 eV[19]. However, the work function of most materials surpasses 3 eV[38], posing a barrier to the practical implementation of thermionic cooling for addressing the overheating problem. A recent experimental demonstration has shown that the height of the Schottky barrier in $WSe_2$-graphene heterostructures can be as low as 0.2 eV[13]. Furthermore, the Schottky barrier height can be readily adjusted by applying a gate voltage to modify the Fermi level of the graphene layer[13]. These two distinctive properties make $WSe_2$-graphene heterostructures a promising platform for investigating phenomena related to thermionic emission.

Here, we demonstrate that the distinctive thermionic emission properties of $WSe_2$-graphene vdW heterostructures can be utilized to enhance heat transfer in vacuum devices. As a proof of concept, we demonstrate that this approach offers a promising solution to the long-standing overheating issue in X-ray tubes. Specifically, we show that the saturated target temperature of a 1200 W X-ray tube can be reduced from around 1200°C to around 490°C by using WSe$_2$-graphene heterostructures-based thermionic cooling compared to copper-based thermionic cooling. Additionally, the performance of thermionic cooling is improved when the Schottky barrier height is adjusted from 0.25 eV to 0.15 eV. Our findings pave the way for the development of high-power X-ray tubes.

Figures 1a and 1b illustrate the work process of an X-ray tube, where an electron beam impinges on a copper (Cu) target, resulting in the emission of X-rays via bremsstrahlung radiation and characteristic radiation. Because of the low energy conversion efficiency of X-ray tubes, only 1% of the input electrical energy is converted into X-rays, while the remaining energy is transformed into heat within the target, leading to an overheating issue[16]. To address the overheating issue, we utilize WSe$_2$-graphene heterostructures-based thermionic cooling to enhance heat transfer. In particular, we incorporated a WSe$_2$-graphene vdW heterostructure (depicted as the brown rectangle in Fig. 1a) at the base of the Cu target. Fig. 1c illustrates the atomic structure of the WSe$_2$-graphene vdW heterostructure. The heat from the target heats up the WSe$_2$-graphene vdW heterostructure. When the thermal energy of the electrons in the vdW heterostructure is sufficiently large to overcome the Schottky barrier height, thermionic emission occurs, leading to a cooling effect as the emitted electrons carry away a portion of the thermal energy. Using the WSe$_2$-graphene vdW heterostructure here is essential, as its low Schottky barrier height allows for the generation of a considerable thermionic current at 500°C. As depicted in Fig. 1d, the current density of thermionic emission from the WSe$_2$-graphene vdW heterostructure at 500°C is approximately 20 orders of magnitude greater than that from a Cu film. The considerable disparity in thermionic emission is attributed to the substantial difference between the work function of Cu (approximately 4.56 eV) and the Schottky barrier height (work function) of the WSe$_2$-graphene vdW heterostructure (around 0.2 eV).

In an X-ray tube with thermionic cooling (Fig. 1a and 1b), the heat of the target is mainly dissipated via three methods, namely, thermal conduction, thermal radiation and thermionic cooling. In accordance with Fourier's law of thermal conduction, the heat dissipated through thermal conduction can be determined by

$$P_c = 2AK(T - T_E), \qquad (1)$$

where $A$ is the surface area of the Cu target, $K$ is the heat transfer coefficient, $T$ is the temperature of the target, $T_E$ is the environmental temperature. The factor of 2 in Eq. (1) accounts for the heat dissipation occurring through both the top and bottom surfaces. In this study, we have assigned the value of $K$ to be $2000\ \text{Wm}^{-2}\text{K}^{-1}$, and we have assumed the target is a $2\times2\times0.05\ \text{cm}^3$ Cu thin film. According to the Stefan-Boltzmann law, the heat dissipated through thermal radiation can be determined by

$$P_r = 2A\epsilon\sigma(T^4 - T_E^4), \qquad (2)$$

where $\epsilon$ is the emissivity, $\sigma = 5.67\times10^{-8}\ \text{J s}^{-1}\text{m}^{-2}\text{K}^{-4}$ is the Stefan-Boltzmann constant. The heat dissipated through WSe$_2$-graphite heterostructure-based thermionic cooling is given by

$$P_{tc} = \frac{A\left[(q\phi + 2k_B T)J(T) - (q\phi + 2k_B T_E)J(T_E)e^{\frac{qV}{k_B T_E}}\right]}{q}, \qquad (2)$$

where $q$ is the charge of an electron, $\phi$ is the Schottky barrier height of the WSe$_2$-graphene vdW heterostructure, $V$ is the external voltage applied to the WSe$_2$-graphene vdW heterostructure and the bottom Cu film (see Fig. 1a), $k_B$ is the Boltzmann constant, and $J(T)$ is the thermionic current density. In our simulations, the value of $V$ is set to 0, as the target temperature is significantly higher than the environmental temperature. For conventional materials such as Cu film, the thermionic current density is determined by the Richardson-Dushman law as follows:

$$J_c = A_c T^2 e^{-\frac{q\phi}{k_B T}}, \qquad (3)$$

where $A_c \approx 1.2\times10^6\ \text{A m}^{-2}\text{K}^{-2}$ is the Richardson constant. For the WSe$_2$-graphene vdW heterostructure, the thermionic current density is given by

$$J_h = \frac{2q}{\pi \tau_{\text{inj}}} \left(\frac{k_B T}{\hbar v_F}\right)^2 \left(\frac{\Phi_0}{k_B T} + 1\right)^2 \exp\left(-\frac{\phi}{k_B T}\right), \qquad (4)$$

where $\tau_{\text{inj}} = 47 \times 10^{-12}$ s is the carrier injection time, $v_F \approx 1 \times 10^6$ m/s is the Fermi velocity of the graphene, $\Phi_0 = 0.54$ eV is the band offset at the WSe$_2$-graphene interface.

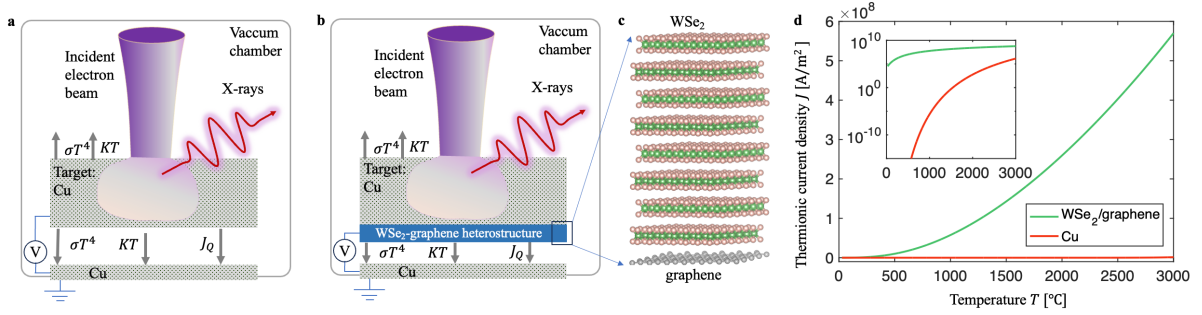

**Fig. 1. Boosting the heat transfer of X-ray tubes through the utilization of WSe₂-graphene heterostructure-based thermionic cooling. (a)** Diagram of an X-ray tube with Copper-based thermionic enhanced heat transfer where incident electron beam impinges the Cu target, resulting in the emission of X-rays. **(b)** Diagram of an X-ray tube with WSe₂-graphene heterostructure-based thermionic cooling. **(c)** The atomic structure of the WSe₂-graphene vdW heterostructure. **(d)** Thermionic current density from Cu and WSe₂-graphene vdW heterostructure.

The heat transfer process in an X-ray tube with thermionic cooling is given by

$$(P_{in} - P_c - P_r - P_{tc})\mathrm{d}t = mC_v \mathrm{d}T, \qquad (5)$$

where $P_{in}$ is the thermal input power of the Cu target, $m$ is the mass of the Cu target, $C_v = 385$ J/(kg K) is the specific heat capacity of Cu, and $\mathrm{d}T$ is the temperature change of the Cu target. Combing Eqs. (1-5), we obtain

$$\mathrm{d}T = \frac{1}{mC_V}[P_{in} - 2AK(T - T_E) - 2A\epsilon(T^4 - T_E^4) - P_{tc}]\mathrm{d}t. \qquad (6)$$

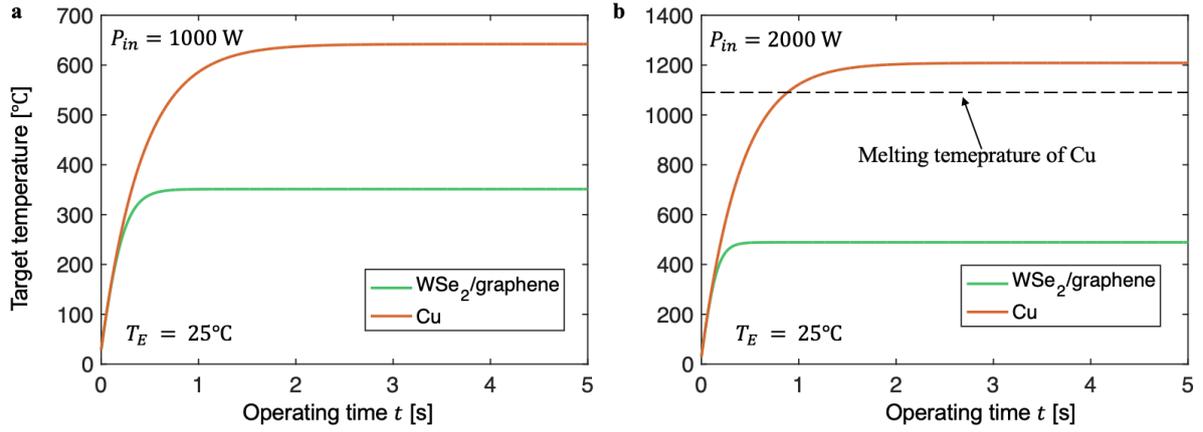

**Fig. 2. Considerably reduce the target temperature of the X-ray tube by employing WSe₂-graphene heterostructure-based thermionic cooling for the following thermal input power levels:** (a) for $P_{in}$ = 1000 W. (b) for $P_{in}$ = 2000 W.

Through the utilization of WSe₂-graphene heterostructure-based thermionic cooling, the target temperature of the X-ray tube is significantly decreased in comparison to that achieved with Cu-based thermionic cooling. In Fig. 2a, it is demonstrated that the utilization of WSe₂-graphene heterostructure-based thermionic cooling reduces the saturated target temperature of the X-ray tube, with thermal input power $P_{in}$ = 1000 W, from 640°C to approximately 350°C, as compared to the employment of Cu-based thermionic cooling. Similarly, for thermal input power $P_{in}$ = 2000 W, the utilization of WSe₂-graphene heterostructure-based thermionic cooling reduces the saturated target temperature of the X-ray tube from 1200°C to approximately 490°C, as compared to the employment of Cu-based thermionic cooling. Since the melting temperature of the target Cu is 1085°C, an X-ray tube utilizing Cu-based thermionic cooling would fail if the thermal input power $P_{in}$ exceeds 2000 W. Conversely, for X-ray tubes employing WSe₂-graphene heterostructure-based thermionic cooling, their saturated target temperature is about 490°C, significantly lower than the melting temperature of 1085°C. The saturated target temperature (490°C) is also significantly lower than the melting temperature of WSe₂ (1200°C) and graphene (3650°C), respectively. This demonstrates the exceptional cooling performance of WSe₂-graphene heterostructure-based thermionic cooling.

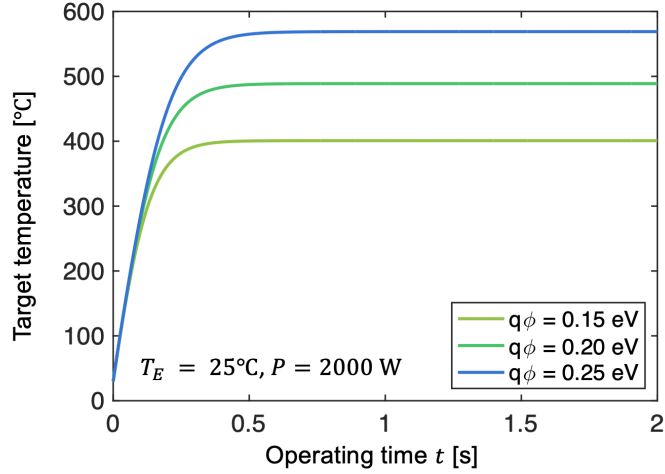

**Fig. 3.** Enhancing the heat transfer performance of WSe$_2$-graphene heterostructure-based thermionic cooling by reducing the Schottky barrier height formed in the WSe$_2$-graphene heterostructures.

The attractiveness of employing WSe$_2$-graphene thermionic cooling to improve heat transfer in X-ray tubes is further enhanced by the ability to easily adjust the Schottky barrier height in WSe$_2$-graphene heterostructures by applying a gate voltage, thereby modifying the Fermi level of the graphene layer[13]. Fig. 3 shows the target temperature of an X-ray tube with WSe$_2$-graphene thermionic cooling is reduced from 570°C to 400°C by reducing the value of Schottky barrier height from 0.25 eV to 0.15 eV.

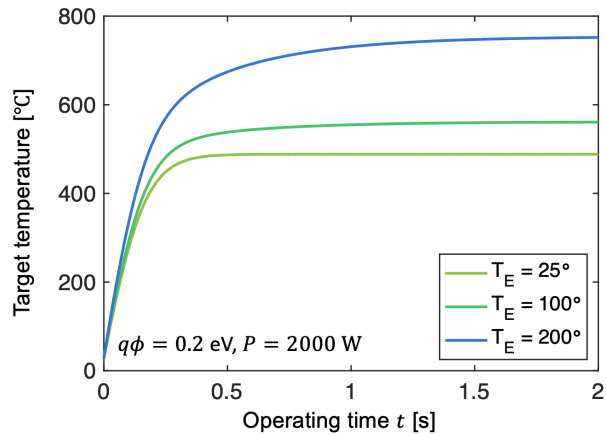

**Fig. 4.** Enhancing the heat transfer performance of WSe$_2$-graphene heterostructure-based thermionic cooling by reducing the environmental temperature.

Typically, the target temperature of an X-ray tube is also influenced by the environmental temperature. In Fig. 4, it is illustrated that the saturated target temperature of an X-ray tube with WSe$_2$-graphene thermionic cooling rises from 488°C to 750°C as the environmental temperature increases from 25°C to 200°C. Therefore, the target temperature of an X-ray tube can be decreased by water-cooling or air-cooling the environment surrounding the X-ray tube. While it is possible to directly cool the target of an X-ray tube using water or air, this method poses considerably greater challenges than cooling the surrounding environment with water or air, given that the target operates within a vacuum environment.

We have proposed the utilization of $WSe_2$-graphene-based thermionic cooling to address the long-standing issue of overheating in X-ray tubes. Our method is highly complementary to other existing methods such as water-cooling technology and rotatable target design. Here, our aim to provide an innovative solution to the overheating issue in X-ray tubes, rather than to compete with existing cooling methods. Although we have showcased the promising applications of $WSe_2$-graphene-based thermionic cooling in X-ray tubes, further experimental investigations are necessary for this method. In our study, we have constrained ourselves to a scenario where the dimensions of the copper target are similar to those of the electron beam. This enables us to assume a uniform target temperature, facilitating the demonstration of the potential application of vdW heterostructure-based thermionic cooling in X-ray tubes without becoming entangled in complex geometry issues. We would like to point out that Cu-based thermionic cooling has negligible impact on the heat transfer of an X-ray tube due to the large work function of Cu. In this context, the study of Cu-based thermionic cooling serves as a benchmark for the performance of $WSe_2$-graphene-based thermionic cooling.

In summary, we have demonstrated the promising application of $WSe_2$-graphene-based thermionic cooling in addressing the long-standing issue of overheating in X-ray tubes. We have shown that the saturated target temperature of a 2000 W X-ray can be reduced from around 1200°C to 490°C by using $WSe_2$-graphene-based thermionic cooling compared to Cu-based thermionic cooling. Furthermore, we have demonstrated that the cooling performance can be improved either by lowering the Schottky barrier height of the $WSe_2$-graphene heterostructure or by reducing the environmental temperature of the X-ray tube. The adoption of $WSe_2$-graphene-based thermionic cooling to tackle the enduring problem of overheating in X-ray tubes is motivated by the experimental demonstration of the Schottky barrier height of the $WSe_2$-graphene heterostructure being as low as 0.2 eV.

This work was supported by National Natural Science Foundation of China (61921002 and 92163204).

**Conflict of Interest**

The authors have no conflicts to disclose.

**DATA AVAILABILITY**

The data that support the findings of this study are available from the corresponding author upon reasonable request.